\documentclass[10pt,final,twocolumn,fleqn,leqno]{IEEEtran}
\title{Hypothesis Elimination in Kleene Semirings\\(Extended Abstract)}
\author{Ernie Cohen \\
(ernie.cohen@microsoft.com)}
\usepackage{abbrev}
\usepackage{latexsym}
\usepackage[fleqn]{amsmath}
\def\Ax{\mathsf{A}}
\def\KA{\mbox{KS}}
\def\Top{\top}

\def\Space{\vspace{10pt}}
\begin{abbreviate}{Math}{:!(=<>/|.}{:!(=<>/|.}
  \x .. {.}
  \x .   {.}
  \x => {\ImpliesSym}
  \x === {\doteq}
  \x ==  {\sim}
  \x =-= {\sim}
  \x =-  {=-}
  \x =   {=}
  \x <<  {\langle}
  \x <=>  {\IffSym}
  \x <=  {\ImpliedSym}
  \x <|  {\{}
  \x <A> {\langle A \rangle}
  \x <A  {<A}
  \x <B> {\langle B \rangle}
  \x <B  {<B}
  \x <f> {\langle f \rangle}
  \x <f  {<f}
  \x <   {\leq}
  \x >>  {\rangle}
  \x >   {\geq}
  \x /|  {\ \AndSym\ }
  \x /   {/}
  \x !>  {\leadsto}
  \x !   {\overline}
  \x (E: {(\exists}
  \x (E  {(E}
  \x (A: {(\forall}
  \x (A  {(A}
  \x (   {(}
  \x |-> {\LeadsToSym}
  \x |-  {\vdash}
  \x |>  {\}}
  \x |   {\vert}
  \x :A  {\leadsto}
  \x :E  {\EnsuresSym}
  \x :U  {\UnlessSym}
  \x :L  {\LeadsToSym}
  \x :S  {\mbox{\bf stable}}
  \x :o: {\circ}
  \x :o  {:o}
  \x :M  {\downarrow}
  \x :D  {\uparrow}
  \x ::  {\mbox{{\bf in}}}
  \x :w: {\InfSym\hspace{-1pt}}
  \x :w  {^\omega}
  \x :u: {\circ}
  \x :u  {^\circ}
  \x :s: {^*.}
  \x :s  {^*}
  \x :   {:}
\end{abbreviate}
\Math
\def\InfSym{\star\:}

\def\False{\mathit{false}}

\def\Mop{\mathsf{op}}

\newenvironment{LProof}{
\begin{math} \begin{array}{l@{\hspace{5mm}}l@{\hspace{5mm}\{}l@{\}}}}{\end{array}\end{math}
}
\def\comment#1{}

\def\AndSym{\wedge}
\def\/{\OrSym}
\def\OrSym{\vee}

\def\ImpliesSym{\Rightarrow}
\def\ImpliedSym{\Leftarrow}
\def\IffSym{\Leftrightarrow}

\def\Def#1{{\em #1}}
\def\Kwd#1{{\mbox{\bf #1}}}

\def\While{\Kwd{while}}

\def\Do{\Kwd{do}}

\def\Ref#1{(\ref{#1})}
\def\Thm#1#2{\begin{equation}#1\end{equation} 
        \nopagebreak  #2  \nopagebreak\noindent $\Box$\par }
\def\Fm#1{ \begin{equation}\begin{aligned} #1 \end{aligned} \end{equation}} 

\def\Defof#1{\mbox{def}\  #1}
\def\So{\ \mbox{so }}

\def\Hyp{\mbox{(hyp)}}
\def\IndHyp{\mbox{(ind hyp)}}
\def\Below{\mbox{(below)}}

\def\SInd{\mbox{(* ind)}}

\begin{document}
\maketitle
\begin{abstract}

A \Def{Kleene Semiring} (KS) is an algebraic structure satisfying the axioms of Kleene 
algebra, minus the annihilation axioms $x..0 = 0 = 0..x$. 
We show that, like Kleene algebra (KA), KS admits efficient elimination of various kinds of
equational hypotheses, in particular Hoare formulas ($x=0$). 
Our method is purely proof-theoretic, and can be used to 
eliminate Horn hypotheses in any suitable Horn-equational theory. Moreover, it gives a simple
condition under which hypotheses eliminations can be combined.
\end{abstract}

\section{Introduction}
Kleene algebra (KA) \cite{KA} and its descendants, such as
Kleene algebra with tests (KAT) \cite{KAT} and omega algebra \cite{OA}
have proved useful in reasoning about programs and program transformations,
e.g. theorems about concurrency control \cite{OA}, static analysis \cite{staticAnalysis}, and compiler optimization \cite{compilers}.
These algebras well-suited to such applications for several reasons.
The operators of Kleene algebra (the regular expression operators
$0$,$1$,$+$,$..$, and $:s$) correspond naturally to program operators (the miracle, skip, 
nondeterministic choice, sequential composition, and finite repetition).
KA is easy to teach and to use; its equational theory 
is just the equational theory of regular expressions, and its formulation of induction is 
particularly simple (e.g., no well-founded sets). KA is particularly well-suited to 
program reasoning requiring commutativity arguments, because arbitrary terms (not just tests)
can be used as inductive hypotheses \cite{OA}. Using tests, KA can faithfully encode
most things one wants to do in PSPACE (e.g., automata constructions), 
including arguments that are awkward in alternative formalisms (e.g., 
a 10 page TLA proof of the reduction theorem in \cite{pretendingatomicity} shrinks to half a page
\cite{OA}). Finally, the equational theory of these algebras is computationally tractable (PSPACE-complete \cite{KA,KATcomplexity}),
in contrast to alternatives such as relational algebra.

Most interesting applications of these algebras require reasoning in the presence of
additional equational hypotheses giving the required properties of the program
fragments. For example, if $p$ and $q$ are tests, the equation 
$p..x..q=0$  represents the Hoare triple $\{p\}\
x\ \{\neg q\}$, and in omega algebra, $(p..x):w = 0$ expresses termination of
the program ``$\While\ p\ \Do\ x$''. Following Kozen, we call such
equations \Def{Hoare formulas}.

An important property of these algebras is that Hoare hypotheses
can be efficiently eliminated using the following theorem, first
proved in \cite{Hyps}: for any ground terms $x$,$y$, and $z$, 
\[ (x=0 |- y=z) <=> (|- f(y)=f(z)) \]
where $f(u) = \Top..x..\Top + u$,  ``$..$'' is the product operator (i.e., sequential composition),
and $\Top$ is the maximal element of
the algebra (or $\Sigma:s$, where $\Sigma$ is the sum of all letters appearing in
$x$,$y$,or $z$). This shows that these algebras remain PSPACE-complete
even of we allow Hoare hypotheses. This flavor of hypothesis elimination has been extended to
other kinds of hypotheses, such as $x<1$ where
$x$ doesn't contain the product operator \cite{Hyps}, $p..a = p$ where
$p$ is a test and $a$ is atomic \cite{Hardin02onthe},
and elimination of combinations of hypotheses \cite{Hardin}.

One of the limitations of these algebras is that the annihilation axioms $0..x = 0 = x..0$
are problematic when we want to reason about total correctness or specifications.
For example, if $x$ is a nonterminating program, we would expect $x..0$ to be equal to $x$ rather than $0$.
Because of this, several KA-like algebras keep $0..x = 0$ but omit $x..0 = 0$, e.g. \cite{DRA,CC,Moller}.
The axiom $0..x=0$ is problematic if $x$ represents a specification of a function with precondition
$\False$ and some non-$\False$ postcondition. We would like to extend 
hypothesis elimination techniques to these weaker algebras.

In this paper, we give a general technique for eliminating a set of Horn-equational hypotheses from a Horn-equational
theory. In contrast to previous hypothesis elimination techniques, which required constructing explicit algebras,
our method is purely proof-theoretic, and so can be used to eliminate appropriate classes of hypotheses 
from any theory under suitable conditions. It also yields a simple condition under which we can combine such eliminations.

Our main result is that Hoare hypotheses can be eliminated from 
\Def{Kleene Semirings} (KS), whose axioms are Kozen's axioms for KA without the annihilation axioms.
Thus, we can replace the annihilation axioms with arbitrary Hoare axioms, 
while keeping the decision procedure for equality in PSPACE.

\subsection{Notation}
In this paper, an ``algebra'' is given by an operator signature (including the set of constants) and a set of (ground) 
Horn-equational axioms on this signature. Thus, the usual presentation of an algebra as 
a set of axioms, with the variables of each universally quantified over elements of the algebra, will be tacitly treated instead
as a set of axiom schemas, where these variables are actually metavariables ranging over terms; we can take this
liberty because we are concerned in this paper only with the ground theory. We will also tacitly take the equality rules as an axiom scheme, and so consider them explicit axioms of
the algebra, i.e. for every operator $\Mop$ of the algebra, and tuples of terms $u$ and $v$, we have an implicit
axiom $u=v => \Mop(u) = \Mop(v)$ (and analogously for the reflexivity and transitivity axioms of equality).
This will allow us to talk about proofs without having to special-case equality.

The algebra under consideration will be determined by the context.
The word ``constant'' means a nullary function of the algebra, ``operator'' means a non-nullary function of the algebra
``term'' means a term of the algebra, ``equation'' means an equation between terms,
and ``formula'' means a Horn formula whose literals are equations.
Except when indicated otherwise, $a$, $b$, $c$, $x$, $y$, and $z$ are metavariables
ranging over terms, $u$ and $v$ are metavariables ranging over tuples of
terms,  $H|-F$ (where $H$ is a set of formulas and $F$ is a formula) means that the conclusion
of $F$ is provable in the algebra whose axioms are those of the algebra along with the 
formulas of $H$ and the hypotheses of $F$, and $H|-C$ (where $C$ is a set of formulas)
means $H|-F$ for every formula $F$ in $C$.

Except when explicitly indicated, all identifiers are single letters, and are either metavariables
representing terms or (meta)functions from  terms to  terms.
Juxtaposition of identifiers always denotes function application
(right associative, e.g., $psfx = p(s(f(x))))$). 
Function application is given precedence higher than the operators of the algebra, 
(e.g. $pa:s = (p(a)):s$). 
We extend the metafunctions to  
tuples of terms, equations, formulas, and sets of formulas by distributing it through
tuples, Boolean connectives, and equalities: 
\begin{eqnarray*}
f(\langle x,y,\dots\rangle) &=& \langle fx, fy, \dots \rangle\\
f((/|i: E_i) => E) &\equiv& ((/|i: f(E_i)) => f(E))\\
f(x=y) &\equiv& (fx=fy)
\end{eqnarray*}

\section{Hypothesis Elimination}
Here we give a general method for hypothesis elimination
in Horn-equational theories. Fix an algebra, let $H$ be a set formulas,
and let $F$ range over sets of formulas, and $E$ range over equations.
To eliminate $H$, we define a suitable \Def{elimination function}
 $f$, for which we establish
\Fm{\label{result} (A: E,F : (H, F |- E) <=> (fF |- fE))}
We prove  \Ref{result} as follows:\Space\\
\begin{LProof}
H, F |- E &=>& \Ref{induc}\\
fF |- fE &=>& |-\\
H, fF |- fE &=>& \Ref{fxx}\\
H, F |- E
\end{LProof}
\Space\\
We are  left with the obligations
\Fm{\label{induc} (A: E,F: (H, F |- E) => (fF |- fE))}
\Fm{\label{fxx}  (A: x: H |- fx=x)}

We prove \Ref{induc} by induction on the proof of $ H,F|-E $. 
Each step of the proof is an axiom of the algebra or
a formula in $H$ or $E$.  The case of
a formula of $F$ is trivial, since we have $fF$ as a hypothesis. For the
cases where the step is an axiom $\Ax$ of the algebra or 
a formula of $H$, we need to show
\Fm{\label{hg} |- f(\Ax)}
\Fm{\label{hg1} |- f(H)}

This leaves as proof obligations \Ref{fxx}, \Ref{hg}, and \Ref{hg1}.
For each case of $H$, we will define a suitable $f$ and show that it satisfies
these obligations.

\subsection{Eliminating multiple hypotheses}
The approach above already allows simultaneous elimination of multiple hypotheses (since $H$ can contain any number of formulas),
but requires sharing a single elimination function $f$. There are two ways to eliminate hypotheses in stages.

Suppose we want to eliminate two sets of hypotheses, $H$ and $H'$, that we can eliminate in some algebra 
using elimination functions $f$ and $f'$ respectively.
One possibility is to use $f$ to reduce $H, H' |- F$ to $fH' |- fF$. However, in general we might not
be able to eliminate $fH'$, even though we could eliminate $H'$.

As an alternative, we can first add $H'$ to the algebra (without changing $F$), eliminate $H$
(in the new algebra), and finally eliminate $H'$ from $f(F)$ in the original algebra. 
The addition of $H'$ to the algebra in the first step introduces a new proof obligation
\Fm{H' |- f(H')}
and the resulting elimination function is $f' \circ f$ (where $f$ is the function applied first).

As an example of this, suppose that the formulas of $H'$ are all equations (without hypotheses),
and that $f$ is a function of the form $fx = x + t$ for some term $t$; this was the form
of the elimination function for hypotheses of the form $a=0$ in Kleene algebra, where $t = \Top.a.\Top$ \cite{Hyps}. 
Then the proof obligation for an equation
$e1=e2$ in $H'$ reduces to $e1=e2 |- e1+t = e2+t$, which follows immediately from equality reasoning. 
This gives a trivial proof that elimination of $a=0$
in Kleene  algebra can be combined with the elimination of other equational hypotheses, which was previously proved in \cite{Hardin}
using a more complex argument.

An obvious generalization is that in any theory, if a set of hypotheses can be eliminated with an elimination function that is polymorphic 
in its argument, then the hypotheses can be eliminated alongside any set of eliminatable equational hypotheses.

\section{Kleene Semirings}
For the rest of the paper, we work in the theory of Kleene semirings,
the axioms of which are simply Kozen's axioms for Kleene algebra
\cite{KA} with the annihilation axioms removed; as usual, $x<y$ abbreviates $x+y=y$:
\[\begin{array}{rcl@{}rcll}
(x+y)+z &=& x+(y+z)\\
x+y &=& y+x\\                        
x+x &=& x                         \\
0 + x &=& x\\                   
x..(y..z)&=& (x..y)..z    \\        
1..x = x..1 &=& x                \\
x..(y+z) &=& x..y + x..z          \\
(x+y)..z &=& x..z + y..z \\
x:s &=& 1 + x + x:s:x:s\\
x..y \leq x &\Rightarrow& x..y:s = x & \SInd \\
x..y \leq y &\Rightarrow& x:s:y = y & \SInd \\
\end{array}\]
The equational theory of KS is, like KA, PSPACE-complete. Moreover,
the equational theory of KS, restricted to terms not mentioning $0$,
is the same as the similarly restricted equational theory of KA.

An initial model of KS can be constructed as follows. (We give an informal construction here;
a more precise construction is given later as a corollary of the hypothesis elimination theorem
for hypotheses of the form $a=0$.) 
Define a \Def{closed} language over an alphabet that includes the symbol $0$ to be a language $L$ such that
 (1) $0 \in L$, and (2) for all strings $r,s,t$ (each possibly empty) such that $r..s..t \in L$, $s..0..t \in L$. 
Define the closure of a language to be the smallest closed language that contains it. 
Interpret each operator of KS as in KA, but operating on closed languages and closing the result.
For example, if $b$ and $c$ are symbols, the the language denoted by the expression
$b..c$ is the the set of strings generated (treating $0$ as an ordinary symbol) from the regular expression
$0:s..(0 + 0..c + b..0 + b..0:s..c)..0:s$. 

An example of a relational model of KS is the following. 
Terms denote binary relations on a set (representing states) with a distinguished element $\bot$ 
(representing nontermination), where each relation maps relates input $\bot$ to an output iff that output is $\bot$. The operators
are interpreted as in the relational model of KA, as are the constants, except for $0$ which is the identity relation restricted to $\bot$. Note that this model satisfies $0..x = 0$, but not $x..0 = 0$.

\noMath
\section{Eliminating $a = 0$}
\Math
In this section, we present our main result, the elimination of equations of the form
$a=0$. 

The absence of the annihilation axioms makes the definition of a suitable elimination function
much more complex than that of  the elimination function used for Kleene algebra \cite{Hyps}.
With the annihilation axioms, the assumption $a=0$ can be viewed as
saying that we are in a modified language model where
terms denote sets of strings not containing superstrings of strings of
$a$.  (Operators in this model behave as usual, then remove
superstrings of $a$ from the result.) This model is isomorphic to one
where terms represent languages that include {\em all} superstrings of
$a$, hence the definition $f(x)=\Top.a.\Top+x$, where $\Top = \Sigma:s$, 
where $\Sigma$ is the sum of all symbols in the alphabet.
But in the absence of the annihilation axioms, this construction does not work, because
we would be unable to prove \Ref{hg1}. 

Without the annihilation axioms, we can no longer imagine working in a simple
string model. Instead, we imagine working in an ordered string model, where string $s$ ``refines'' string
$t$ iff $t$ can be transformed into $s$ by a sequence of improvement
steps, where a string is improved by replacing an arbitrary substring
with an arbitrary string of $a$ (or the string $0$). Terms now denote
nonempty sets of strings closed under refinement (i.e.  if a set
contains $t$ and $s$ refines $t$, then the set contains $s$).

In this model, we can think of $f$ as the closure operator (i.e.,
$fx$ computes the language of all strings that refine strings of
$x$).  The problem is how to define $f$ as a function from terms
to terms.  Because substrings of strings from $a$ added in
improvement steps can themselves be rewritten by later improvement
steps, the key is to define a term $m$ that gives an explicit formula
for $fa$. $fx$ itself can then be defined as the set of strings obtainable
by breaking up a string of $x$ into a finite set of substrings (some of which might be empty)
and replacing some of these substrings with strings from $m$.

Formally, for any term $x$,  define $px$ (the ``prefixes'' of $x$)
and $sx$ (the ``suffixes'' of $x$) as follows ($c$ ranges over all
constant symbols, including $0$ and $1$):
\[
\begin{array}{lll@{\hspace{1cm}}lll}
pc &=& 1+c  & sc &=& 1+c\\
p(x+y)  &=& px + py          & s(x+y) &=& sx + sy\\
p(x..y) &=& px + x..py       & s(x..y) &=& sy + sx..y\\
p(x:s) &=& x:s:px              & s(x:s) &=& sx..x:s\\
\end{array} \]
The theorems we need regarding these functions are proved in the appendix.
Note that we cannot simply claim obvious properties like these because they hold
for ordinary languages, since we are effectively in an ordered language model.

We next define the terms $l$ and $m$ as follows:
\begin{eqnarray*}
l &=& pa:s:a..sa:s \\
m &=& l..(psa..l):s
\end{eqnarray*}
Intuitively, $m$ consists of all of the strings obtainable by starting with a string 
of $a$ and repeatedly replacing an arbitrary substring with a string of $a$ (or $0$). 

Finally, we define the function $f$ as follows, by induction on the term structure of
its argument:
\begin{eqnarray*}
fc &=& (1+m)..(c+m)..(1+m) \mbox{ for constant }c\\
f(x+y) &=& fx + fy\\
f(x..y) &=& fx..fy + pfx..m..sfy\\
f(x:s) &=& gx:s\\
gx&=& fx + pfx..m..(psfx..m):s:sfx
\end{eqnarray*}
Intuitively, the strings of $fx$ are strings of $x$, chopped into substrings (some empty),
with $m$ substituted for some of the substrings.  The following proofs show that this $f$ 
satisfies the formulas  \Ref{fxx} (\Ref{fxxproof} and \Ref{mfproof} below), 
\Ref{hg} (\Ref{hgproof} below), and \Ref{hg1} (\Ref{l0} below).

The following facts are proved about $p$ and $s$ (all  by induction on $x$, 
except for the last which is proved by induction on the derivation of
$|-x=y$). 
\begin{align*}
\Ref{1xpx} &\ 1+x < px&\\
\Ref{ppx} &\ ppx = px&\\
\Ref{pssp} &\ psx =spx&\\
\Ref{pxz} &\ px.0 = x.0&\\
\Ref{pAlg} &\ (|- x=y) => (|- px=py)&
\end{align*}

The following properties are proved by direct calculation, using the
definitions of $l$ and $m$ and the above properties of $p$ and $s$:
\begin{align*}
\Ref{mpsam}&\  m.psa.m < m&\\
\Ref{pl} &\ pl = pa:s + l.psa&\\
\Ref{plm} &\ pl.m = m&\\
\Ref{pmm}  &\ pm..m = m&\\
\Ref{mpsm} &\ m..psm..m < m&
\end{align*}

The following properties are proved by induction on $x$:
\begin{align*}
\Ref{xmf} &\ x+m < f.x&\\
\Ref{fact} &\ x<psm /| fx < x + px.m.sx => pfx.m < px.m&\\
\Ref{main} &\ x<psm => fx < x + px..m..sx&
\end{align*}
The remaining properties are proved by direct calculation:
\begin{align*}
\Ref{l0} &\ x<m => fx=m&\\
\Ref{pg} &\ pgx = pfx.(pm + (m.psfx):s)&\\
\Ref{pgxmsgx} &\ pgx.m.sgx < gx&
\end{align*}

\section{Properties of $p$ and $s$}

\Thm{\label{1xpx} 1+x < px}{
\mbox{Proof by induction on $x$:}\\
\begin{LProof}
x=c:\\
px &=&\Defof{x}\\
pc &=&\Defof{p}\\
1+c &=& \Defof{x}\\
1+x
\end{LProof}
\Space\\
\begin{LProof}
x=y+z:\\
px &=& \Defof{x}\\
p(y+z) &=& \Defof{p}\\
py+pz &>&py>1+y,\ pz>1+z\  \IndHyp\\
(1+y)+(1+z) &=&\KA\\
1+(y+z) &=& \Defof{x}\\
1+x
\end{LProof}
\Space\\
\begin{LProof}
x=y..z:\\
px &=&\Defof{x}\\
p(y..z) &=&\Defof{p}\\
py+y.pz &>&[y>1+y,\ pz>1+z\  \IndHyp\\
(1+y)+y.(1+z) &>& \KA\\
1+y.z &=& \Defof{x}\\
1+x
\end{LProof}
\Space\\
\begin{LProof}
x=y:s:\\
px &=& \Defof{x}\\
p(y:s) &=& \Defof{p}\\
y:s:py &>& py > 1+y\ \IndHyp\\
y:s:(1+y) &>& \KA\\
1 + y:s &=& \Defof{x}\\
1+x
\end{LProof}
\Space\\
}

\Thm{\label{ppx} ppx = px}{
\mbox{Proof by induction on $x$:}
\Space\\
\begin{LProof}
x=c:\\
ppx &=&\Defof{x}\\
ppc &=&\Defof{p}\\
p(1+c) &=& \Defof{p}\\
(1+1)+(1+c) &=& \KA\\
1+c &=& \Defof{p}\\
pc &=& \Defof{x}\\
pc
\end{LProof}
\Space\\
\begin{LProof}
x=y+z:\\
ppx &=& \Defof{x}\\
pp(y+z) &=& \Defof{p}\\
p(py+pz) &=&\Defof{p}\\
ppy+ppz &=&ppy=py,\ ppz=pz\ \IndHyp\\
py+pz &=&\Defof{p}\\
p(y+z) &=&\Defof{x}\\
px
\end{LProof}
\Space\\
\begin{LProof}
x=y..z:\\
ppx &=& \Defof{x}\\
pp(y..z) &=& \Defof{p}\\
p(py+y.pz) &=& \Defof{p}\\
ppy+ py+y.ppz &=&ppy=py,\ ppz=pz\ \IndHyp\\
py+y.pz &=& \Defof{p}\\
p(y.z) &=&\Defof{x}\\
px
\end{LProof}
\Space\\
\begin{LProof}
x=y:s:\\
ppx &=& \Defof{x}\\
pp(y:s) &=& \Defof{p}\\
p(y:s:py) &=& \Defof{p}\\
p(y:s) + y:s:ppy &=&ppy=py,\ \IndHyp\\
p(y:s) + y:s:py &=&\Defof{p}\\
p(y:s)
\end{LProof}
\Space\\}

\Thm{\label{pssp} psx = spx}{
\mbox{Proof by induction on $x$:}\Space\\
\begin{LProof}
x=c:\\
psx &=& \Defof{x}\\
psc &=& \Defof{s}\\
p(1+c) &=& \Defof{p}\\
1+c &=& \Defof{s}\\
s(1+c) &=&\Defof{p}\\
sp(c) &=&\Defof{x}\\
spx
\end{LProof}
\Space\\
\begin{LProof}
x=y+z:\\
psx &=& \Defof{x}\\
ps(y+z) &=& \Defof{s}\\
p(sy+sz) &=& \Defof{p}\\
psy+psz &=& psy=spy,\ psz=spz\ \IndHyp\\
spy+spz &=& \Defof{s}\\
s(py+pz) &=& \Defof{p}\\
sp(y+z) &=& \Defof{x}\\
spx
\end{LProof}
\Space\\
\begin{LProof}
x=y..z:\\
psx &=& \Defof{x}\\
ps(y..z) &=& \Defof{s}\\
p(sy.z+sz) &=& \Defof{p}\\
psy+sy.pz + psz &=& psy=spy,\   psz=spz\ \IndHyp\\
spy+sy.pz+spz &=& \Defof{s}\\
s(py+y.pz) &=& \Defof{x}\\
spx
\end{LProof}
\Space\\
\begin{LProof}
x=y:s{:}\\
psx &=& \Defof{x}\\
ps(y:s) &=& \Defof{s}\\
p(sy.y:s) &=& \Defof{p}\\
psy + sy.p(y:s) &=& \Defof{p}\\
psy + sy.y:s:py &=& psy=spy\ \IndHyp\\
spy + sy.y:s:py &=& \Defof{s}\\
spy + s(y:s).py &=& \Defof{s}\\
s(y:s:py) &=& \Defof{p}\\
sp(y:s) &=&\Defof{x}\\
spx
\end{LProof}
\\
}

\Thm{\label{pxz} px.0 = x.0}{
\mbox{Proof by induction on $x$:}\Space\\
\begin{LProof}
x=c:\\
px.0 &=& \Defof{x}\\
pc.0 &=& \Defof{p}\\
(1+c).0 &=& \KA\\
1.0+c.0 &=& 1.0=0;\ 0 + c.0 =c.0\\
c.0
\end{LProof}
\Space\\
\begin{LProof}
x=y+z:\\
px.0 &=& \Defof{x}\\
(py+pz).0 &=& \KA\\
py.0+pz.0 &=& py.0=y.0,\ pz.0=z.0\ \IndHyp\\
y.0+z+0 &=& \KA\\
(y+z).0 &=& \Defof{x}\\
x.0
\end{LProof}
\Space\\
\begin{LProof}
x=y..z:\\
px.0 &=& \Defof{x}\\
p(y..z).0 &=& \Defof{p}\\
(py+y.pz).0 &=& \KA\\
py.0 + y.pz.0 &=& py.0=y.0,\ pz.0=z.0\ \IndHyp\\
y.0 + y.z.0 &=& \KA\\
y(0+z.0) &=& \KA\\
y.z.0 &=& \Defof{x}\\
x.0
\end{LProof}
\Space\\
\begin{LProof}
x=y:s{:}\\
px.0 &=& \Defof{x}\\
p(y:s).0 &=& \Defof{p}\\
y:s:py.0 &=& py.0=y.0\ \IndHyp\\
y:s:y.0 &<& y:s:y < y:s \ \KA\\
y:s:0 &=& \Defof{x}\\
x.0 &<& x<px\ \Ref{1xpx}\\
px.0
\end{LProof}
\Space\\
Since the first and last terms are equal, the first and sixth terms are equal.
\\}

\Thm{\label{pShort} 
\begin{array}{rcl}
 px:s:px &=& px\\
 p(x:s:y) &=& x:s:(px+py)\\
 p(px:s:y) &=& px:s:py\\
 p(px.y) &=& px.py
\end{array}
}{
\begin{LProof}
px:s:px &<& \KA\\
px:s &<& 1<px \ \Ref{1xpx}\\
px:s:px
\end{LProof}
\Space\\
\begin{LProof}
p(x:s:y) &=&\Defof{p}\\
p(x:s) + x:s:py &=& \Defof{p}\\
x:s:px + x:s:py &=&\KA\\
x:s:(px+py)
\end{LProof}
\Space\\
\begin{LProof}
p(px:s:y) &=&\mbox{proof above} \\
px:s:(ppx+ py) &=&ppx=px\ \Ref{ppx}\\
px:s:(px+ py) &=&px:s:px = px:s\mbox{ (proof above)}\\
px:s:(1+ py) &=&1<py \ \Ref{1xpx}\\
px:s:py
\end{LProof}
\Space\\
\begin{LProof}
p(px.y) &=& \Defof{p}\\
ppx + px.py &=& \Ref{ppx}\\
px + px.py &=& 1<py\ \Ref{1xpx},\ \So px<px.py\\
px.py
\end{LProof}
\Space\\
}

\Thm{\label{pAlg}  (|- x=y) => (|-px=py) }{
\mbox{Proof by induction on the proof of $x=y$.}\\
\begin{LProof}
(x+y)+z = x+(y+z):\\
p((x+y)+z) &=&\Defof{p}\\
(px + py)+pz &=& \KA\\
px + (py+pz) &=& \Defof{p}\\
p(x+(y+z))
\end{LProof}
\Space\\
\begin{LProof}
x+y=y+x:\\
p(x+y) &=& \Defof{p}\\
px+py &=& \KA\\
py+px &=& \Defof{p}\\
p(y+x)
\end{LProof}
\Space\\
\begin{LProof}
x+x=x :\\
p(x+x) &=& \Defof{p}\\
px+px &=& \KA\\
px 
\end{LProof}
\Space\\
\begin{LProof}
0+x=x:\\
p(0+x) &=& \Defof{p}\\
p0+px &=& \Defof{p}\\
1+0 + px &=& 1< px\ \Ref{1xpx}\\
px 
\end{LProof}
\Space\\
\begin{LProof}
x.(y.z)=(x.y).z :\\
p(x.(y.z)) &=& \Defof{p}\\
px + x.(py+y.pz) &=& \KA \\
px+x.py + x.y.pz &=& \Defof{p}\\
p((x.y).z)
\end{LProof}
\Space\\
\begin{LProof}
1..x=x:\\
p(1..x) &=&\Defof{p}\\
p1 + 1..px &=&\Defof{p}\\
1 + px &=& \Defof{p}\\
px
\end{LProof}
\Space\\
\begin{LProof}
x..1=x:\\
p(x..1) &=& \Defof{p}\\
px + x.p1 &=& \Defof{p}\\
px + x &=& x<px \ \Ref{1xpx}\\
px
\end{LProof}
\Space\\
\begin{LProof}
x..(y+z) = x..y+x..z:\\
p(x..(y+z)) &=&\Defof{p}\\ 
px + x..p(y+z) &=& \Defof{p}\\
px + x..(py+pz) &=&\KA\\
px + x..py + x..pz &=&\KA\\ 
(px + x..py) + (px + x..pz) &=&\Defof{p}\\
p(x..y) + p(x..z) &=&\Defof{p}\\
p(x..y + x..z)
\end{LProof}
\Space\\
\begin{LProof}
(x+y)..z = x..z+y..z:\\
p((x+y)..z)&=&\Defof{p}\\
p(x+y) + (x+y)..pz &=& \Defof{p}\\
px+py+x..pz+y..pz  &=&\KA\\
(px+x..pz) + (py+y..pz) &=&\Defof{p}\\
p(x..z) + p(y..z) &=&\Defof{p}\\
p(x..z+y..z)
\end{LProof}
\Space\\
\begin{LProof}
x:s=1+x+x:s:x:s :\\
p(1+x+x:s:x:s) &=& \Defof{p}\\
1+px + p(x:s:x:s) &=&\Ref{pShort}\\
1+px + x:s:(px + x:s:px) &=&px < x:s:px\ \KA\\
1+px + x:s:x:s:px &=& x:s:x:s = x:s \ \KA\\
1+px + x:s:px &=& 1+px < x:s:px\ \KA\\
x:s:px &=&\Defof{p}\\
p(x:s)
\end{LProof}
\Space\\
\mbox{$x..y<x => x..y:s=x$:}\\ 
\mbox{Suppose $x..y<x$.}\\
\mbox{Then by the induction hypothesis, $p(x..y)<px$, so} 
\begin{LProof}
p(x..y:s) &=& \Defof{p}\\
px+ x..y:s:py &=& x..y:s = x \\ 
px + x..py &=& \Defof{p} \\
p(x..y) &<& \IndHyp\\
px &<& \Defof{p}\\
p(x..y:s)
\end{LProof}
\Space\\
For the induction axiom $x..y<y => x:s:y=y$, suppose $x..y<y$. Then 
by the induction hypothesis, $p(x..y)= px+x..py<py$, so
\Space\\
\begin{LProof} 
p(x:s:y) &=& \Ref{pShort}\\
x:s:(px + py) &=& px < py \ \Hyp\\
x:s:py &=& x..py<py\ \IndHyp\\
&& \So x:s:py=py \SInd\\
py
\end{LProof}
\Space\\
}

\subsection{Properties of $m$}

\Thm{\label{mpsam} m.psa.m < m}{
\begin{LProof}
m.psa.m &=&\Defof{m}\\
l.(psa.l):s:psa.l.(psa.l):s &<&(psa.l).(psa.l):s< (psa.l):s\\
l.(psa.l):s:(psa.l):s &=&(psa.l):s:(psa.l):s = (psa.l):s\\
l.(psa.l):s &=&\Defof{m}\\
m
\end{LProof}
\Space\\}

\Thm{\label{pl} pl = pa:s + l.psa}{
\begin{LProof}
pl &=& \Defof{l}\\
p(pa:s:a..sa:s) &=& \Ref{pShort}\\
pa:s:p(a..sa:s) &=& \Defof{p}\\
pa:s:(pa + a.sa:s:psa) &=&  pa:s:pa = pa:s \Ref{pShort}\\
pa:s + pa:s:a.sa:s:psa) &=& pa:s:a.sa:s = l\ \Defof{l}\\
pa:s + l.psa 
\end{LProof}
\Space\\}
\Thm{\label{plm} pl.m = m}{
\begin{LProof}
pl.m &=& pl=pa:s + l.psa\ \Ref{pl}\\
(pa:s + l.psa)..m&=& \Defof{m}\\
(pa:s + l.psa).l.(psa.l):s &<&psa.l.(psa.l):s < (psa.l):s\\
(pa:s + 1).l.(psa.l):s &<&1<pa:s\\
pa:s:l.(psa.l):s &=& \Defof{l}\\
pa:s:(pa:s.psa.sa:s).(psa.l):s &=& pa:s:pa:s = pa:s\ \KA\\
pa:s.psa.sa:s.(psa.l):s &=& \Defof{l}\\
l.(psa.l):s &=& \Defof{m}\\
m &<& 1<pl \Ref{1xpx}\\
pl.m
\end{LProof}
\Space\\}

\Thm{\label{pmm} pm.m = m}{
\begin{LProof}
pm.m &=& \Defof{m}\\
p(l..(psa..l):s).m &=& \Defof{p}\\
(pl + l..(psa..l):s:p(psa..l)).m &=& \Defof{m}\\
(pl + m.p(psa..l)).m &=& p(psa.l)=psa.pl\ \Ref{pShort}\\
(pl + m.psa.pl).m &=&pl.m=m\ \Ref{plm}\\
m + m.psa.m &=&m.psa.m < m\ \Ref{mpsam}\\
m
\end{LProof}
\Space\\}

\Thm{\label{mpsm} m.psm.m < m}{
\begin{LProof}
m.psm.m &<& m.psm < pm+m.psm \\
(pm+m.psm)..m	&=&\Defof{p}\\
p(m.sm).m &=& m.sm = m\ \Ref{pmm}\\
pm.m &=& pm.m = m\ \mbox{dual of }\Ref{pmm}\\
m 
\end{LProof}
\Space\\
}

\subsection{Properties of $f$ and $g$} \label{fProps}

\Thm{\label{xmf} x+m < fx}{
\mbox{Proof by induction on $x$:}
\Space\\
\begin{LProof}
x=c:\\
fx &=& \Defof{x}\\ 
fc &=& \Defof{f}\\
(1+m).(c+m).(1+m) &>& \KA\\
c+m &=& \Defof{x}\\
x+m
\end{LProof}
\Space
\begin{LProof}
x=y+z:\\
fx &=& \Defof{x}\\
f(y+z) &=& \Defof{f}\\
fy+fz &>& y+m<fy\ \IndHyp\\
y+m+fz &>& z+m<fz\ \IndHyp\\
y+m+z+m &=&\KA\\
(y+z)+m &=& \Defof{x}\\
x+m
\end{LProof}
\Space
\begin{LProof}
x=y..z:\\
f(y..z) &=& \Defof{f}\\
fy.fz + pfy.m.sfz &>& 1<pfy,\ 1<sfz\ \Ref{1xpx}\\
fy.fz + m &>& y<fy, z<fz \ \IndHyp\\
y.z+m &=& \Defof{x}\\
x+m
\end{LProof}
\Space
\begin{LProof}
x=y:s{:}\\
f(y:s) &=& \Defof{f,g}\\
(fy + pfy..m..(psfy..m):s:sfy):s &>& 1<(psfy.m):s\\
(fy + pfy..m..sfy):s &>& 1<pfy\ \Ref{1xpx}\\
(fy + m..sfy):s &>&1< sfy\ \Ref{1xpx}\\
(fy + m):s &>& y<fy\ \IndHyp\\
(y+m):s &>& \KA\\
y:s + m &=&\Defof{x}\\
x+m
\end{LProof}
\Space\\}

\Thm{\label{fact}
x<psm /| fx < x + px.m.sx => pfx.m < px.m}{
\begin{LProof}
pfx.m &<& fx < x + px.m.sx\\
p(x + px.m.sx).m &=& \Defof{p},\ \Ref{pShort}\\
(px + px.(pm + m.psx)).m &<& x<psm\ \Hyp\\
(px + px.(pm + m.psm)).m &<& m.psm.m < m \ \Ref{mpsm}\\
(px + px.(pm + 1)).m &=& pm.m = m\ \Ref{pmm}\\
px.m 
\end{LProof}
\Space\\}

\Thm{\label{main} x<psm => fx < x + px..m..sx}{
\mbox{Proof by induction on $x$: assuming $x<psm$,}
\Space\\
\begin{LProof}
x=c:\\
fx &=& \Defof{f}\\
(1+m)..(c+m)..(1+m) &=& \KA;\ m..m<m \Ref{pmm}\\
c+c..m + m..c + m..c..m + m &<& c<px,\ c<sx\ \Ref{1xpx}\\
&&\ c<psm\ \Hyp\\
x+px..m + m..sx + m..psm..m &<& m..psm..m < m \ \Ref{mpsm}\\
x+px..m + m..sx + m &<& 1<px,\ 1<sx\ \Ref{1xpx}\\
x+px..m..sx
\end{LProof}
\Space\\
\begin{LProof}
x=y+z:\\
fx &=&\Defof{f}\\
fy + fz &<&y<x<psm\\
&&z<x<psm\ \Hyp\\
&&\IndHyp\\
y+py.m.sy + z + pz.m.sz &<& \KA\\
(y+z) + (py+pz)..m..(sy+sz) &=& \Defof{p}\\
(y+z) + p(p+z)..m..s(y+z) &=& \Defof{x}\\
x + px.m..sx
\end{LProof}
\Space
\begin{LProof}
x=y..z:\\
fx &=&\Defof{f}\\
fy..fz + pfy..m..sfz &<&y<px<ppsm=psm\\
&&\IndHyp,\ \Ref{fact}\\
fy..fz + py..m..sfz &<&z<sx<spsm=psm\\
&&\IndHyp\\
&&\mbox{dual of }\Ref{fact}\\
fy..fz + py..m..sz &<&\IndHyp\\
(y+py.m.sy)\\
\ \ ..(z+pz.m.sz)\\
\ \ + py..m..sz &<&py+pz+y.pz \\ 
  && <px\ sy+sz+sy.z\\
  && <sx\\
y.z+px.m.sx \\
\ \ + px.m.sy.pz.m.sx &<&y..z=x\\
&&sy.pz<psx<psm\\
&& \Ref{mpsm}\\
x+ px.m.sx 
\end{LProof}
\Space
\begin{LProof}
x=y:s{:}\mbox{ let }z=py.m.sy;\mbox{ then }\\
fx &=&\Defof{f}\\
(fy + pfy.m.(psfy.m):s:sfy):s &<&psfy < psm\\
  &&m.psm.m < m\\
(fy + pfy.m.sfy):s &<&y<x<psm;\\&& \Ref{fact}\\
(fy + z):s &<&y<x<psm;\\&& \IndHyp\\
(y + z):s &<&\KA\\
y:s:(z.y:s):s &<&y:s < y:s + z.y:s\\
(y:s + z.y:s).(z.y:s):s &<&
  (y:s+ z.y:s).(z.y:s) \\
	&& < y:s + z.y:s \Below\\
	&&\SInd\\
y:s + z.y:s &=& \Defof{z}\\
y:s + px.m.sy.y:s &=& sy.y:s = s(y:s)\ \Defof{s}\\
x+px.m.s(y:s) &=&\Defof{x}\\
x+px.m.sx
\end{LProof}
\Space\\
\begin{LProof}
(y:s+ px.m.sy.y:s).z.y:s &=& y:s:py = px\\
	&& sy.y:s:py = psx\\
(px + px.m.psx).m.sy.y:s &=& x<psm\Hyp \\
  &&\So psx < pspsm = psm\\
  &&\Ref{pssp}, \Ref{ppx}\\
(px + px.m.psm).m.sy.y:s &<&m.psm.m < m \ \Ref{mpsm}\\
px.m.sy.y:s &<& 0 < y:s\\
y:s + px.m.sy.y:s
\end{LProof}
\Space\\}

\Thm{\label{l0} x<m => fx=m}{
\begin{LProof}
fx &<& x<m<psm;\ \Ref{main}\\
px.m.sx &<& x<m;\mbox{ monotonicity of }p\\
pm.m.sm &=&\Ref{pmm}\\
m &<&\Ref{xmf}\\
fx
\end{LProof}
\Space\\
}

\Thm{\label{pg} pgx = pfx.(pm + (m.psfx):s)}{
\mbox{Let $r=psfx.m$ and $t=psfx.pm$; then}
\begin{LProof}
pgx &=& \Defof{g}\\
p(fx + r..r:s.sfx) &=&\Defof{p}\\
pfx + p(pfx.m.r:s.sfx) &=&\Ref{pShort}\\
pfx + pfx.p(m.r:s.sfx) &=&\Defof{p}\\
pfx + pfx.(pm+ m.p(r:s.sfx) &=&1<pm\ \Ref{1xpx}\\
pfx.(pm+ m.p(r:s.sfx) &=&\Ref{pShort}\\
pfx.(pm+ m.r:s.(pr + psfx)) &=&\Ref{pShort}\\
pfx.(pm+ m.r:s.(t + psfx)) &=&\KA\\
pfx.(pm+ r:s.m.(t + psfx)) &=&\KA\\
pfx.(pm + (m.psfx):s) 
\end{LProof}
\Space\\}

\Thm{\label{pgxmsgx} pgx.m.sgx < gx}{
\begin{LProof}
pgx.m.sgx &=&\Ref{pg}\\
pfx.(pm + (m.psfx):s)\\
\ \  ..m.(sm + (psfx.m):s).sfx &=&pm.m=m \ \Ref{pmm}\\
pfx.(m.psfx):s.m.(psfx.m):s:sfx &<&\KA\\
pfx.m.(psfx.m):s:sfx&<& \Defof{g}\\
gx
\end{LProof}
\Space\\}

\Thm{\label{pfxm} pfx.m < fx}{
\mbox{Proof by induction on the structure of $x$:}
\Space
\begin{LProof}
x=c:\\
pfx.m &=&\Defof{x}\\
pfc.m &=&\Defof{f}\\
p((1+m).(c+m).(1+m)).m &=&\Defof{p}\\
(1+pm\\
\ \ +m..(1+c+pm+m.(1+pm))).m &=&\KA\\
(pm+m.(c+pm+m.pm)).m &=&pm.m =m\\
(1+m.(c+1+m)).m &<&m.m<m\\
(1+m.(c+1)).m &<&\Defof{f}\\
fc&=&\Defof{x}\\
fx
\end{LProof}
\Space\\
\begin{LProof}
x=y+z:\\
pfx.m &=&\Defof{x}\\
pf(y+z).m &=&\Defof{f}\\
p(fy+fz).m &=&\Defof{p}\\
(pfy+pfz).m &<&pfy.m<fy\\
 && pfz.m < fz\ \IndHyp\\
fy+fz &=&\Defof{f}\\
f(y+z) &=&\Defof{x}\\
fx
\end{LProof}
\Space\\
$x=y..z$: \mbox{let $r=pfy.m.psfz$; then}\\
\begin{LProof}
pfx.m &=&\Defof{x}\\
pf(y..z).m &=&\Defof{f}\\
p(fy.fz + pfy.m.sfz).m &=&\Defof{p}\\
(pfy+fy.pfz + ppfy\\
\ \ + pfy.pm + r).m &=&ppfy=fpy\ \Ref{ppx}\\
(pfy+fy.pfz \\ \ \ + pfy.pm + r).m &=&pm.m=m\\
(pfy+fy.pfz + r).m &=& \KA\\
pfy.m + fy.pfz.m + r.m &=& psfz.m = spfz.m \\
	&& < s(pfz.m)\\
pfy.m + fy.pfz.m \\
\ \ + pfy.m.s(pfz.m) &=& pfy.m<fy\\
	&& pfz.m < fz\\
	&&\IndHyp\\
pfy.m + fy.fz + pfy.m.sfz &=& 1<sfz\ \Ref{1xpx}\\
	&&pfy.m<pfy.m.sfz\\
fy.fz + pfy.m.sfz &=& \Defof{f}\\
fx
\end{LProof}
\Space\\
\begin{LProof}
x=y:s{:}\\
pfx.m &=& \Defof{x}\\
pf(y:s).m &=&\Defof{f}\\
p(gy:s).m &=&\Defof{p}\\
gy:s.pgy.m &=& 1<sgy\ \Ref{1xpx}\\
gy:s.pgy.m.sgy &<& \Ref{pgxmsgx}\\
gy:s.gy &<&\KA\\
gy:s &=&\Defof{f}\\
f(y:s) &=&\Defof{x}\\
fx
\end{LProof}
\Space\\
}
\noMath
\subsection{$f$ preserves axioms}
\Math

\Thm{\label{hgproof}  |- f(\Ax)}{
\mbox{Proof by case analysis of $\Ax$:} 
\Space\\
\begin{LProof}
(x+y)+z = x+(y+z):\\
f((x+y)+z) &=& \Defof{f}\\
f(x+y)+fz &=& \Defof{f}\\
fx+fy+fz &=& \Defof{f}\\
fx+f(y+z) &=& \Defof{f}\\
f(x+(y+z))
\end{LProof}
\Space\\
\begin{LProof}
x+y=y+x:\\
f(x+y) &=& \Defof{f}\\
fx+fy &=& \KA\\
fy+fx &=& \Defof{f}\\
f(y+x)
\end{LProof}
\Space\\
\begin{LProof}
x=x=x:\\
f(x+x) &=& \Defof{f}\\
fx+fx &=& \KA\\
fx
\end{LProof}
\Space\\
\begin{LProof}
0+x=x:\\
f(0+x) &=& \Defof{f}\\
f0 + fx &=& \Defof{f}\\
m + fx &=& m<fx \ \Ref{xmf}\\
fx
\end{LProof}
\Space\\
\begin{LProof}
x.(y+z)=x.y+x.z:\\
f(x.(y+z)) &=&\Defof{f}\\
fx.(fy+fz) + pfx.m.(sfy+sfz) &=&\KA\\
fx.fy + pfx.m.sfy + fx.fz + pfx.m.sfz &=&\Defof{f}\\
f(x.y) + f(x.z) &=&\Defof{f}\\
f(x.y+x.z)
\end{LProof}
\Space\\
\begin{LProof}
1.x=x:\\
f(1..x) &=&\Defof{f}\\
f1.fx + pf1.m.sfx &=& \Defof{f}\\
(1+m).fx+(1+pm).m.sfx &<& pm.m = m \ \Ref{pmm}\\
fx + m.fx+m.sfx  &=&fx<sfx\ \Ref{1xpx}\\
	&&\So m.fx < m.sfx\\
fx + m.sfx &<& m.sfx < fx\ \Ref{pfxm}\\
fx
\end{LProof}
\Space\\
\begin{LProof}
x+x=x:\\
f(x+x) &=& \Defof{f}\\
fx+fx &=& \KA\\
fx 
\end{LProof}
\Space\\
\begin{LProof}
a=0:\\
fa &=& \Ref{l0}\\
m &=&\Ref{l0}\\
f0
\end{LProof}
\Space\\
$(x.y).z=x.(y.z)$: \mbox{let $r = pfx.m$; then}
\Space\\
\begin{LProof}
f((x.y).z) &=&\Defof{f}\\
f(x.y).fz + pf(x.y).m.sfz &=&\Defof{f}\\
(fx.fy + r.sfy).fz + (pfx+fx.pfy\\
\ \ + pfx.pm + r.psfy).m.sfz &=&pm.m=m\\
(fx.fy + r.sfy).fz + (pfx+fx.pfy\\
\ \  + r.psfy).m.sfz&=&\KA\\
fx.(fy.fz + pfy.m.sfz) \\
\ \ + r.(sfy.fz + sfz \\
 \ \ + psfy.m.sfz)&=&m=m.sm\\
fx.(fy.fz + pfy.m.sfz) \\
\ \ + r.(sfy.fz + pm.sfz + sfz \\
\ \ +  psfy.m.sfz)&=&\Defof{f}\\
fx.f(y.z) + r.sf(y.z) &=&\Defof{f}\\
f(x.(y.z))
\end{LProof}
\Space\\
$x:s=1+x+x:s:x:s$\\\mbox{let $r = p(gx:s).m.s(gx:s)$; then}
\\
\begin{LProof}
f(1+x+x:s:x:s) &=&\Defof{f}\\
f1 + fx + (gx:s).(gx:s) + r&=&\Defof{f,p,s}\\
m+ 1 + fx + gx:s.gx:s +r&=&gx:s:gx:s=gx:s\\
	&&1 + m + fx < gx:s\\
gx:s + gx:s.pgx.m.sgx.gx:s&<&pgx.m.sgx < gx\ \Ref{pgxmsgx}\\
gx:s &=&\Defof{f}\\
f(x:s)
\end{LProof}
\Space\\
$x..y<x => x..y:s=x$:\\
\mbox{assume $f(x..y)<fx$.}\\
\mbox{then $fx.fy + pfx.m.sfy = f(x..y) <fx $, so}\\
\begin{LProof}
f(x..y:s) &=&\Defof{f,s}\\
fx.gy:s + pfx.m.sgy.gy:s &<& pfx.m.sgy < fx \Below\\
fx.gy:s &=& fx.gy < fx \ \Below;\ \SInd\\
fx &<& \KA \\
fx.gy:s + pfx.m.sgy.gy:s	&=&\Defof{f}\\
f(x..y:s)
\end{LProof}
\Space\\
\begin{LProof}
fx.gy &=&gy < fy+fx.pfy.m.sgy\\&&  \Defof{gy}\\
fx.fy + fx.pfy.m.sgy &<&fx.pfy < p(fx.fy) < pfx\\
fx.fy + pfx.m.sgy &<&pfx.m.sgy < fx \ \Below\\
fx.fy + fx &<&fx.fy<fx\\
		&&\Hyp\\
fx
\end{LProof}
\Space\\
\begin{LProof}
pfx.m.sgy &=&\mbox{dual of }\Ref{pg}\\
pfx.m.(sm + (psfy.m):s).sfy &=& m.sm = m\ \Ref{pmm}\\
pfx.m.(psfy.m):s:sfy &=& pfx.m.(psfy.m) \\
	&&< pfx.m\ \Below\\&& \SInd\\
pfx.m.sfy &<& \Hyp\\
fx
\end{LProof}
\Space\\
\begin{LProof}
pfx.m.psfy.m &<&\Defof{p}\\
p(pfx.m.sfy).m &<&pfx.m.sfy< fx\ \Hyp\\
pfx.m
\end{LProof}
}

\Thm{\label{fxxproof} a=0 |- f(x) = x}{
\mbox{Proof: induction on $x$ (using $a=0 |- m=0$)}:
\begin{LProof}
x=c:\\
fx &=& \Defof{x}\\
fc &=&\Defof{f}\\
(1+0).(0+c).(1+0) &=&\KA\\
c 
\end{LProof}
\Space\\
\begin{LProof}
x=y+z:\\
fx &=&\Defof{x}\\
f(y+z) &=& \Defof{f}\\
fy+fz &=& fy=y,\ fz=z\ \IndHyp\\
y+z
\end{LProof}
\Space\\
\begin{LProof}
x=y..z:\\
fx &=& \Defof{x}\\
f(y..z) &=& \Defof{f}\\
 fy.fz + pfy.0.sfz  &=& \Ref{pxz} \\
fy.fz +fy.0.fz &=&0<1\\
fy.fz &=& fy=y,\ fz=z\ \IndHyp\\
y.z = x
\end{LProof}
\Space\\
\begin{LProof}
x=y:s{:}\\
fx &=& \Defof{x}\\
f(y:s) &=& \Defof{f},g\\
(fy+ pfy.0.(psfy.0):s:sfy):s &<& pfy.0=fy.0\ \Ref{pxz} \\
(fy+ fy.0.(psfy.0):s:sfy):s &<& \KA \\
(fy+ fy.(0..psfy):s..0..sfy):s &<& 0..sfy=0..fy\ \Ref{pxz} \\
(fy+ fy..(0..fy):s..0..fy):s &=&fy=y\ \IndHyp\\
(y+ y.(0..y):s..y):s &<& \KA\\
y:s &=& \Defof{x}\\
x
\end{LProof}
}

\Thm{\label{mfproof} (|-f(u)=f(v)) => (|-f(\Mop(u)=f(\Mop(v))))}{
\mbox{Case analysis on $\Mop$ (using \Ref{pAlg}):}
\begin{LProof}
\Mop= +:\\
f(u_0+u_1) &=&\Defof{f}\\
fu_0+fu_1 &=&fu_0=fv_0,\ fu_1=fv_1\ \Hyp\\
fv_0+fv_1 &=& \Defof{f}\\
f(v_0+v_1)
\end{LProof}
\Space\\
\begin{LProof}
\Mop= .:\\
f(u_0.u_1) &=&\Defof{f}\\
fu_0.fu_1  + pfu_0.m.sfu_1 &=& \Hyp\\
fv_0.fv_1  + pfv_0.m.sfv_1 &=&\Defof{f}\\
f(u_0.u_1) 
\end{LProof}
\Space\\
\begin{LProof}
\Mop= :s{:}\\
f(u:s) &=& \Defof{f,g}\\
(fu + pfu.m.(psfu.m):s:sfu):s &=&\Hyp\\
(fv + pfv.m.(psfv.m):s:sfv):s &=& \Defof{f,g}\\
f(v:s) 
\end{LProof}
}

\bibliography{bib}

\end{document}